\documentclass[prl,aps,reprint, amsmath,amssymb,superscriptaddress,showkeys]{revtex4-1}
\usepackage[utf8]{inputenc} 
\usepackage{graphicx} 
\usepackage{hyperref} 
\usepackage{multirow} 
\usepackage{tabularx} 
\usepackage{color} 
\usepackage{textcomp} 
\usepackage{amsmath} 
\usepackage{amssymb} 
\usepackage{amsfonts} 
\usepackage{amsxtra} 
\usepackage{wasysym} 
\usepackage{isomath} 
\usepackage{mathtools} 
\usepackage{txfonts} 
\usepackage{upgreek} 
\usepackage{enumerate} 
\usepackage{tensor} 
\usepackage{pifont} 
\usepackage{soul} 

\usepackage{siunitx}

\definecolor{color-1}{rgb}{0,0,1}
\begin{document}
\title{Irreversible hardening of a colloidal gel under shear: the smart response of natural rubber latex gels.}

\author{Guilherme de Oliveira Reis}

\affiliation{UMR IATE, INRA, CIRAD, Universit\'{e} de Montpellier, Montpellier SupAgro, 2, pl Viala, 34090 Montpellier, France}

\author{Thomas Gibaud} 

\affiliation{Univ Lyon, Ens de Lyon, Laboratoire de Physique, CNRS, F-69342 Lyon, France.} 

\author{Brice Saint-Michel}

\affiliation{Univ Lyon, Ens de Lyon, Laboratoire de Physique, CNRS, F-69342 Lyon, France.} 

\author{S\'{e}bastien Manneville}

\affiliation{Univ Lyon, Ens de Lyon, Laboratoire de Physique, CNRS, F-69342 Lyon, France.} 

\author{Mathieu Leocmach} 

\affiliation{Institut Lumière Matière, UMR5306 Université Claude Bernard Lyon 1 - CNRS, Université de Lyon, 69622 Villeurbanne, France}

\author{Laurent Vaysse} 

\affiliation{UMR IATE, INRA, CIRAD, Universit\'{e} de Montpellier, Montpellier SupAgro, 2, pl Viala, 34090 Montpellier, France}

\author{Fr\'{e}d\'{e}ric Bonfils} 

\affiliation{UMR IATE, INRA, CIRAD, Universit\'{e} de Montpellier, Montpellier SupAgro, 2, pl Viala, 34090 Montpellier, France}

\author{Christian Sanchez} 

\affiliation{UMR IATE, INRA, CIRAD, Universit\'{e} de Montpellier, Montpellier SupAgro, 2, pl Viala, 34090 Montpellier, France}

\author{Paul Menut} \email{paul.menut@supagro.fr}

\affiliation{UMR IATE, INRA, CIRAD, Universit\'{e} de Montpellier, Montpellier SupAgro, 2, pl Viala, 34090 Montpellier, France}

\affiliation{Ing\'{e}nierie Proc\'{e}d\'{e}s Aliments, AgroParisTech, INRA, Universit\'{e} Paris-Saclay, 91300, Massy, France.}

\altaffiliation{Postal address~: UMR GENIAL - AgroParisTech/INRA, 1 avenue des Olympiades, 91744 MASSY cedex, FRANCE\\
Tel~: 0 (33) 1 69 93 50 19\\
Fax~: 0 (33) 1 69 93 50 05}

\begin{abstract}
Natural rubber is obtained by processing natural rubber latex, a liquid colloidal suspension that rapidly gels after exudation from the tree. We prepared such gels by acidification, in a large range of particle volume fractions, and investigated their rheological properties. We show that natural rubber latex gels exhibit a unique behavior of irreversible strain hardening: when subjected to a large enough strain, the elastic modulus increases irreversibly. Hardening proceeds over a large range of deformations in such a way that the material maintains an elastic modulus close to, or slightly higher than the imposed shear stress. Local displacements inside the gel are investigated by ultrasound imaging coupled to oscillatory rheometry, together with a Fourier decomposition of the oscillatory response of the material during hardening. Our observations suggest that hardening is associated with irreversible local rearrangements of the fractal structure, which occur homogeneously throughout the sample.  
\end{abstract}

\keywords{Isotropic particles, strain stiffening/hardening, mechano-responsive gels, re-structuring}

\maketitle


\section*{{Introduction}}

Natural Rubber Latex (NRL) is naturally produced by the \textit{Hevea brasiliensis} tree, in laticiferous cells located in the tree bark. This colloidal suspension is released by wounded trees after tapping and rapidly solidifies to form a protective layer against further attacks. The composition of the exudate is complex \cite{Jacob1993}, but the main physical properties are attributed to rubber particles accounting for more than 90\% of the dry mass. Their diameters range from a few hundred nanometers to a few micrometers. Rubber particles mostly consist of a poly(cis-1,4-isoprene) rubber core surrounded by a membrane of phospholipids \cite{Sansatsadeekul_2011,Cornish_1999} and proteins \cite{dennis1989rubber}. Their surface is negatively charged at native pH (pH ${\approx}$ 7), which ensures net repulsion and therefore latex stability within the tree. After exudation, the pH rapidly decreases under bacterial acidification \cite{Salomez_2018}; the particle surface charge density drops and the suspension is consequently destabilized. Particles aggregate in fractal-type clusters, which eventually form a gel \cite{de_Oliveira_Reis_2015}. 

To produce Natural Rubber (NR), NRL gels are compressed and dried so that the residual water and its soluble non-isoprene content are extracted. During this process, rubber particles lose their individuality and the material is described as a structured network of cis-poly-isoprene chains loosely cross-linked with non-isoprene compounds (mainly lipids and proteins) \cite{Wu_2017,Tanaka_2009}. The cross-link density of these polymeric materials is further increased during a process called vulcanization. Vulcanized NRs can sustain up to 800\% deformation before rupture, and their stress-strain relationship is highly nonlinear. Above a critical strain, whose value depends on the preparation process, but is generally above 300\%, vulcanized NR exhibits strain stiffening: its apparent elastic modulus increases rapidly with strain. This behavior, also known as strain hardening, is attributed to strain-induced crystallization \cite{Karino_2007},\cite{Trabelsi_2003}. It is highly sensitive to the supramolecular organization in the natural product and is called “green strength” \cite{Amnuaypornsri2009}. NR exhibits such unique thermo-mechanical properties that more than 40\% of all the rubber used worldwide is, even today, of natural origin, despite fundamental drawbacks, such as its variability. Understanding the structure-properties relationship in such materials is still a challenge, and requires an understanding of how the structure is formed at the different stages of NRL processing into NR. 

We focus here on the mechanical properties of NRL gels, which are soft materials in which rubber particles form a percolating network in water. As NRL suspensions are very variable due to agronomic parameters (clone, tapping system), or edaphoclimatic factors (mainly the season) \cite{Galiani_2011}, we work with a commercially available NRL suspension of pre-vulcanized rubber particles, which is stabilized in ammonia and far less variable. In spite of a slightly different composition of the continuous phase, such commercial suspensions behave similarly to the fresh exudate: gelation is observed under similar acidification conditions and the elastic modulus of the gels formed from commercial or native NRL are roughly the same. We induce continuous and controlled acidification of our NRL suspensions in the Couette geometry of a rheometer, in order to form gels that can be mechanically characterized in situ, with no stress history. In previous work focusing on gel formation \cite{de_Oliveira_Reis_2015}, we showed that such gels display strain hardening behavior, which is examined in detail in this article. 

In the following, we first describe the main features of the hardening phenomenon under oscillatory shear. We demonstrate that this behavior is irreversible and occurs over a large range of deformations. We then show that NRL gels display a fractal scaling of their elastic modulus at rest. The spring constant between particles determined from this analysis is compatible with known values of rubber elasticity. In the third section, we show that the hardening amplitude decreases with the particle volume fraction. We further investigate the nonlinear behavior through a Fourier analysis of the strain response and show that hardening is associated with the emergence of odd harmonics, whose strength grows with the imposed stress. These observations are further discussed in the last section, where we propose a simplified picture of the structural reorganization that might lie behind this unique behavior.

\section*{{Materials and Methods}}

\subsection*{Materials} 

Natural rubber latex was obtained from Dalbe, France (ref. 4770002). The volume-based diameter 2\textit{a} of the pre-vulcanized particles is \SI{1}{\micro\metre}, as measured previously by static light scattering \cite{de_Oliveira_Reis_2015}. To remove ammonia, suspensions were first diluted with an equivalent amount of Tris-HCl buffer (pH 8.5, ionic strength = 7 mM). TRIS (purity ${\geq}$99\%, CAS 77-86-1) and HCl (purity 37\%, ${\leq}$1 ppm free chlorine) were obtained from Merck, Germany.  

Then, they were dialyzed for 7 days against the same buffer using a dialysis tube (SpectraPor, cut-off 12-14kDa, diameter 29 mm), before dilution with the buffer to reach the desired volume fraction $\phi_{v}$. Suspensions with $\phi_{v}>$ 0.3 were obtained by direct dilution of the commercial suspension with the buffer, without dialysis, which does not impact the final properties as will be shown in this paper. All the prepared suspensions were stored at 5$^{\circ}$ C and used within three weeks. 

\subsection*{Gelation} 

Glucono-${\updelta}$-lactone (GDL, ref G2164, purity ${\geq}$99\%, Sigma-Aldrich) was used to reduce the pH of the suspension uniformly and gradually. For $\phi_{v}<$ 0.1, a concentration of 1 \%wt GDL set as a standard condition was used to reach a final pH close to 4 in less than 1 h. For $\phi_{v}>$ 0.1, the amount of GDL was increased to reproduce the standard acidification kinetics \cite{de_Oliveira_Reis_2015}. GDL was dispersed in the suspension by stirring for 1 min, before loading into the Couette cell of the rheometer. 

\subsection*{Rheological characterization}

Rheological properties were determined using three stress-controlled rheometers: an AR2000ex (TA Instruments), equipped with a stainless-steel smooth Couette concentric cylinder geometry (inner radius 14 mm, outer radius 15 mm and height 42 mm) and a home-made stainless-steel serrated Couette geometry (same dimensions with saw tooth serrations 1 mm in depth and 1 mm base on each side) and an ARG2 (TA Instruments) equipped with a smooth PMMA (Poly(Methyl MethAcrylate)) Couette geometry (inner radius 23 mm, outer radius 25 mm and height 60 mm). Thanks to their feedback loop, these instruments can be used either in strain- or stress-controlled mode. 

Gelation was first monitored for 4 hours thanks to small-amplitude oscillatory shear at a constant frequency of 1 Hz and a constant strain amplitude of 0.5\%. The results showed no measurable impact of the oscillation amplitude of 0.5\% on the gelation process and on the final gel properties, particularly on the elastic modulus at rest $G’_{0}$. The gel was then characterized with a frequency sweep from 0.01 to 10 Hz at a constant strain of 0.5\%, which was followed by a stress sweep at a constant frequency of 1 Hz, starting from 0.02 Pa up to the fracture of the gel. 

\subsection*{Local rheology} 

We used local oscillatory rheology from echography (LORE) to reconstruct time-resolved local displacements during oscillatory stress experiments \cite{Saint_Michel_2016},\cite{Saint_Michel_2017}. In LORE, rheological measurements are synchronized with high-frequency ultrasonic echography in order to map the local strain response of the NRL gel subject to an oscillatory stress. LORE is performed with the ARG2 (TA Instruments) rheometer equipped with the PMMA Couette geometry of gap $e = \SI{2}{\milli\metre}$. The gels are seeded with almost density matched polystyrene particles (Dynoseeds TS-20 of diameter \SI{20\pm 0.1}{\micro\metre}) to provide ultrasound contrast. Their concentration of 1\% wt yields sufficient ultrasound intensity yet is small enough to prevent multiple scattering and to ensure that the polystyrene particles do not affect the mechanical properties of the sample.

In brief, our high-frequency ultrasonic imaging device relies on a linear array of piezoelectric transducers with a total active length of \SI{32}{\milli\metre} \cite{Saint_Michel_2017,Gallot_2013}. The transducer array is immersed in a water tank surrounding the Couette cell. Short ultrasonic plane pulses with a central frequency of 15 MHz propagate across the gap. These pulses get scattered by the seeding particles within the gel and the backscattered pressure signal is recorded by the transducer array, leading to an “ultrasonic speckle” signal with 128 measurement lines in the vorticity direction $z$ and 640 sampling points in the radial direction $r$. The spatial resolution along the $z$-direction is \SI{250}{\micro\metre} and \SI{100}{\micro\metre} in the $r$-direction with a sampling of 600 images per oscillation period \cite{Saint_Michel_2017}. For each oscillatory stress experiment, a 2D-ultrasonic scan with 20 sequences of five oscillation periods each is recorded. The analysis of the ultrasonic data gives access to the displacement maps ${\Updelta}$ within the Couette cell as a function of time $t$ and of the position $(r,z)$ across the gap, where $r$ is the radial distance to the stator and $z$ the position along the vertical direction. We checked that such displacement maps are always invariant along the $z$-direction. Hence, we focus thereafter on their average over $z$, ${\Updelta}(r,t)$, from which we compute the local strain $\gamma$\textit{\cite{Saint_Michel_2017}}. Under an imposed oscillatory stress, the strain response of the sample may be nonlinear in contrast to the sinusoidal stress input. This results in the presence of harmonics in the Fourier series decomposition of the strain 
$\gamma (t)=\sum _{k}\gamma _{k}\cos\left[2\pi kft+\varphi _{k}\right]$ 
, where $\gamma_{k}$ and $\varphi_{k}$ are the amplitude and phase of the $k^{\mathrm{th}}$ harmonics.

\section*{{Results and discussion}}

\subsection*{NRL gels exhibit irreversible hardening under shear}

\begin{figure*}
\includegraphics[width=\textwidth]{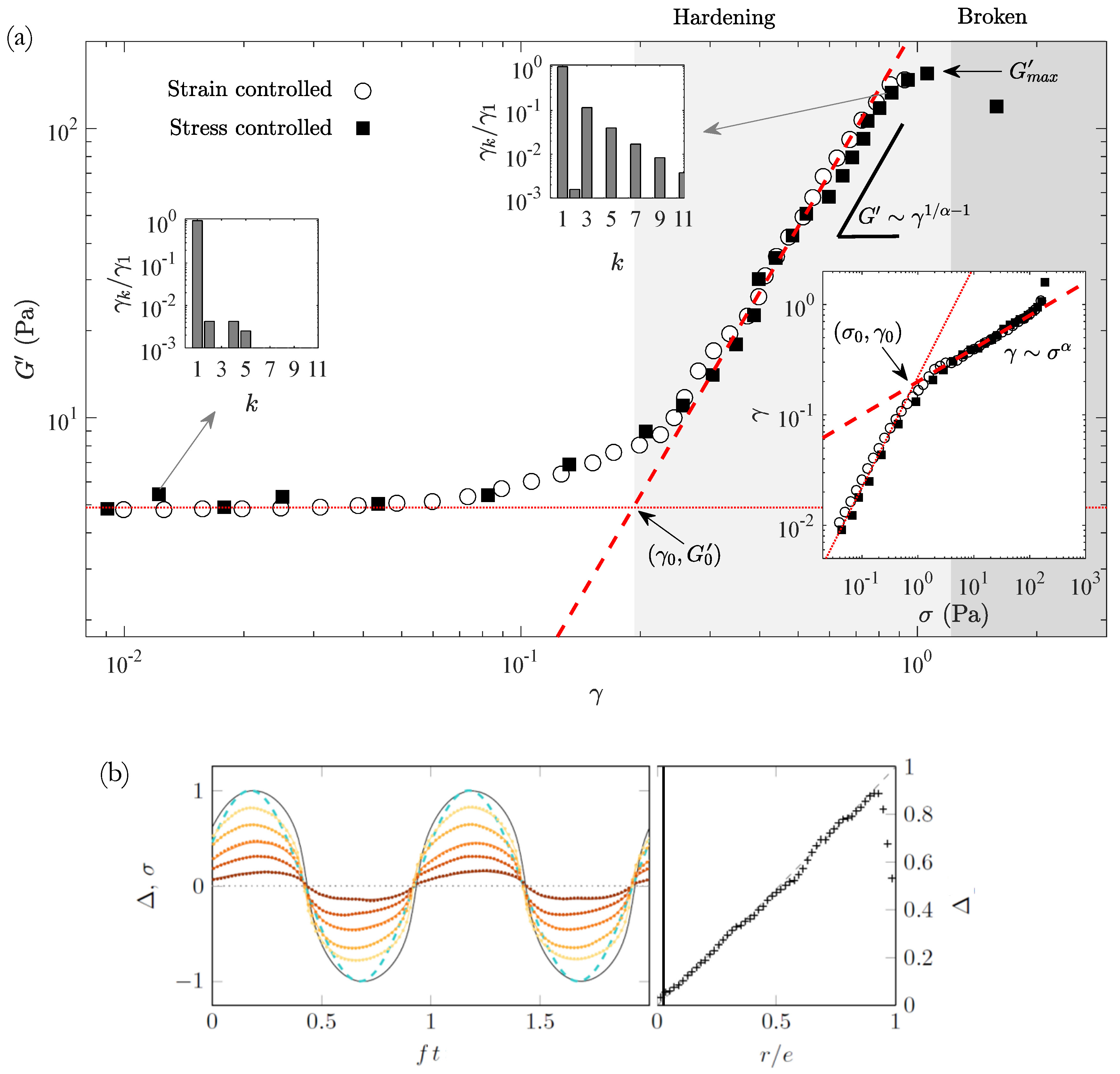}
\caption{(a) Evidence for hardening in NRL gels: change in the elastic modulus, $G’$, under imposed oscillatory strain (open symbols) and under imposed oscillatory stress (filled symbols) for a NRL gel at $\phi_{v} = 0.04$. The continuous line indicates the value $G’_{0}=\SI{5}{\pascal}$ of the elastic modulus at rest and the dashed line is the best power-law fit $G’ \sim \gamma^{1/\alpha-1}$ with $\alpha$=0.32 in the hardening regime, before rupture. Their intersection defines the onset of hardening (${\gamma}_{0}$ ,$G_{0})$. The relative amplitude of the harmonics ${\gamma}_{k}/{\gamma}_{1}$ obtained from the Fourier decomposition of the strain response is shown for two stress-imposed measurements in the linear and in the hardening regimes. Inset: stress-strain relationship for the same data set. (b) local rheology from echography (LORE) in Couette geometry for a stress amplitude $\sigma=\SI{1.8}{\pascal}$ (corresponding to a strain amplitude ${\gamma}=0.23$). Left panel: Local displacements ${\Updelta}(r,t)$ (linear color scale from brown at the stator to yellow at the rotor) and strain ${\gamma}(t)$ recorded by the rheometer (black line) in response to oscillatory stress ${\sigma}(t)$ (blue dashed line) as a function of the normalized time \textit{ft}. All quantities are normalized. Right panel: the amplitude ${\Updelta}_{1}$ of the fundamental Fourier mode of ${\Updelta}(r,t)$ as a function of \textit{r/e} where $r$ is the distance from the stator and $e$ is the gap width. The dashed line shows the theoretical profile for a linear homogeneous strain field in Couette geometry.}
\label{fig:1}
\end{figure*}

When submitted to some mechanical strain, Natural Rubber Latex gels harden. This is seen through oscillatory shear rheology where the material is submitted to oscillations of increasingly large amplitudes in either strain-controlled or stress-controlled experiments. When the amplitude of the imposed oscillations is small, the elastic modulus $G’$ remains independent of the oscillation amplitude. This corresponds to the linear viscoelastic regime. However, for large enough strain amplitudes, typically $\gamma>0.2$, $G’$ steeply increases until material rupture as illustrated for $\phi_{v} =0.04$ in Figure 1. This behavior clearly differs from the strain hardening of dense rubber materials, a phenomenon observed at much higher deformations (${\gamma}>3$) and associated with a transient, and reversible, local crystallization of polymer chains \cite{Karino_2007,Trabelsi_2003}. In the present hardening regime, which is observed on NRL gels in both strain- and stress-controlled modes, the strain-stress relationship is no longer linear: above the onset of hardening, for $\sigma>\sigma_{0}$ and $\gamma>\gamma_0$, one has $\gamma\sim\sigma^\alpha$ with $\alpha\approx0.3$ (inset of Fig.~\ref{fig:1}). Correspondingly, the elastic modulus grows as $G’=\sigma/\gamma\sim\gamma^{1/\alpha-1}\sim\gamma^{2.3}$ at large strains as seen in Fig.~\ref{fig:1}. For the stress-controlled experiments, this corresponds to $G’\sim\sigma^{1-\alpha}\sim\sigma^{0.7}$. Further decomposing of the strain response in a Fourier series shows that strain hardening is in line with the emergence of odd harmonics (insets of Fig.~\ref{fig:1}). Lastly, for $\gamma>1$, the gel fails. 

We conducted LORE experiments to map local displacements ${\Updelta}(r,t)$ in the sample and thereby investigate whether hardening is localized or homogeneously distributed throughout the material. For each imposed stress amplitude, local deformation is recorded during five successive oscillations at $f=\SI{1}{\hertz}$. As shown in Fig.~\ref{fig:1}b (left) for a stress amplitude $\sigma$=1.8 Pa, corresponding to a strain amplitude ${\gamma}=0.23$ at the beginning of the hardening regime, the strain response is distorted. Thus the strain response is clearly nonlinear and contains significant higher-order harmonics. Yet the local strain amplitude ${\Updelta}$ varies linearly across the gap (Fig.~\ref{fig:1}b, right): the bulk material undergoes homogeneous deformation. There is no sign of slip at the walls nor any significant curvature in ${\Updelta}(r)$ that could demonstrate some degree of strain localization, e.g. through localized strain stiffening or strain thinning, over length scales larger than the LORE spatial resolution of about \SI{100}{\micro\metre}. This suggests that the structural reorganization responsible for hardening is localized at scales below \SI{100}{\micro\metre} and is homogeneously distributed over the entire material. In addition, once a stress oscillation is imposed, we do not observe any temporal change in the strain response over the first five oscillations. Hardening therefore occurs within the first oscillation in the bulk material. 

Strain hardening has been reported for colloidal gels obtained by fractal aggregation of nanometer-sized polystyrene beads \cite{Gisler_1999} or globular proteins \cite{Pouzot_2004,Pouzot_2006}. Hardening  has also been observed in biopolymer-based hydrogels \cite{Jaspers_2014,Dobrynin_2011,Licup_2015}. NRL gels however exhibit two very specific properties that we now discuss in more details. 

\begin{figure}
\includegraphics[width=\columnwidth]{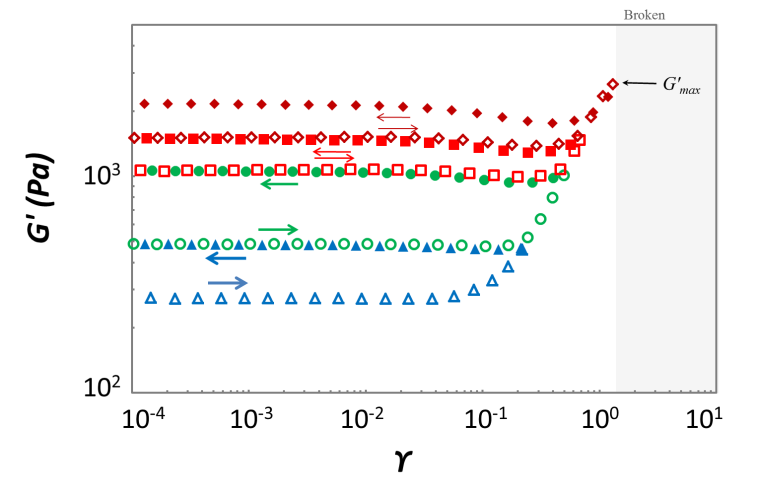} 
\caption{Change in the elastic modulus $G’$ during successive strain sweeps (from blue to red) for a NRL gel with $\phi_{v} = 0.11$. Each cycle consists of (i) a continuous increase in strain amplitude $\gamma$  (empty symbols) up to a predetermined maximum strain $\gamma_\mathrm{max}$, followed (ii) by a continuous decrease in strain amplitude (filled symbols). Each new cycle (from blue to brown), performed on the same sample reaches a higher value of $\gamma_\mathrm{max}$.}
\label{fig:2}
\end{figure}

First, the hardening reported here is \textit{irreversible}: if at a given strain $\gamma_\mathrm{max}$ during material hardening under an upward strain sweep, one decides to decrease the strain amplitude, the material remains “harder” and the elastic modulus $G’$ remains roughly constant whatever the strain amplitude ${\gamma}<\gamma_\mathrm{max}$, as illustrated in Fig.~\ref{fig:2}. Increasing the strain amplitude again and above $\gamma_\mathrm{max}$ is associated with further increase of the elastic modulus, as long as the material is not irreversibly fractured. To our knowledge, such striking irreversible hardening has never been reported before: it suggests that in NRL gels, strain hardening is not associated with a transient increase in ${\sigma}/{\gamma}$, but instead with a permanent increase in the elastic modulus, together with an extension of the material linear regime up to $\gamma_\mathrm{max}$, the maximum strain amplitude reached in the material previous history. Hardening in NRL gels should therefore be associated with some irreversible modifications of the microstructure at the local scale. Note also that during the last cycles in Fig.2, a slight dip is observed in G’ when the material is deformed in a range of strains where it has been previously hardened. This reproducible behavior was observed identically during a further gradual increase or decrease in strain. 

\begin{figure}
\includegraphics[width=\columnwidth]{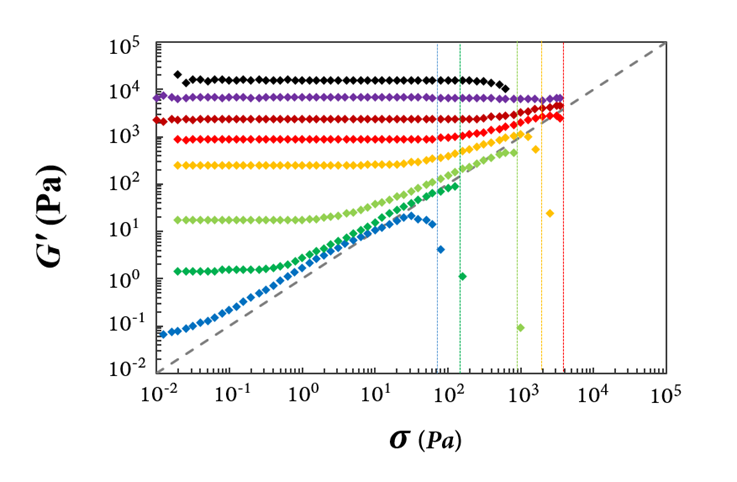}
\caption{Changes in the elastic modulus $G’$ as a function of the imposed stress amplitude $\sigma$ for increasing particle volume fractions: 0.011 (blue), 0.022 (dark green), 0.055 (green), 0.109 (yellow), 0.162 (red), 0.216 (brown), 0.32 (purple) and 0.53 (black). Vertical dashed lines indicate the various stresses at gel rupture. The gray dashed line $G’={\sigma}$ is drawn for comparison. }
\label{fig:3}
\end{figure}

Second, hardening is observed in deformation ranges that are much larger than previously reported for colloidal gels. This allows us to investigate large increases in the stress. In particular, when switching to stress-imposed experiments, we observe in Fig.~\ref{fig:3} that during hardening, the gels elastic modulus remains of the same order of magnitude as the imposed stress amplitude. Thus the material adapts to the stress in order to sustain it as long as fracture does not occur. This provides an elegant way of tuning the elastic modulus of NRL gels at will, simply by applying an oscillatory stress with an amplitude of the order of the target modulus. Of course, this behavior has some limitations: the maximum stress and therefore the elastic modulus that a gel can reach depend on its volume fraction, as seen in Fig.~\ref{fig:3}. The maximum stress at rupture is reached for $\phi_{v}\approx 0.162$, and decreases for $\phi_{v}>0.32$. Below, we first characterize NRL gels behavior at rest, before investigating the hardening behavior in more details.   

\subsection*{Fractal scaling of the elastic modulus at rest} 

We investigated NRL gels for volume fractions ranging from 0.011 to 0.53. As the minimum volume fraction for gel formation is $\phi_{v}\approx 0.01$, and the volume fraction at random closed packing is ${\phi}_{\mathrm{rcp}}\approx 0.715$ \cite{de_Oliveira_Reis_2015}, this roughly covers the full range of volume fractions accessible for gel formation. The elastic modulus at rest spans more than five decades, from \SI{7e-2}{\pascal} to \SI{2.3e4}{\pascal}, and scales with the particle volume fraction $\phi_{v}$ as $G'_{0} \sim\phi_{v}^{3.3}$ (Figure 4). We checked that measurements conducted in geometries with different materials and different roughnesses lead to the exact same results, confirming the absence of wall slip. The power-law behavior of $G'_{0}$ with ${\phi}_{v}$ is reminiscent of fractal gels obtained by colloidal aggregation in the diffusion-limited cluster aggregation regime \cite{Buscall_1988}. 

\begin{figure}
\includegraphics[width=\columnwidth]{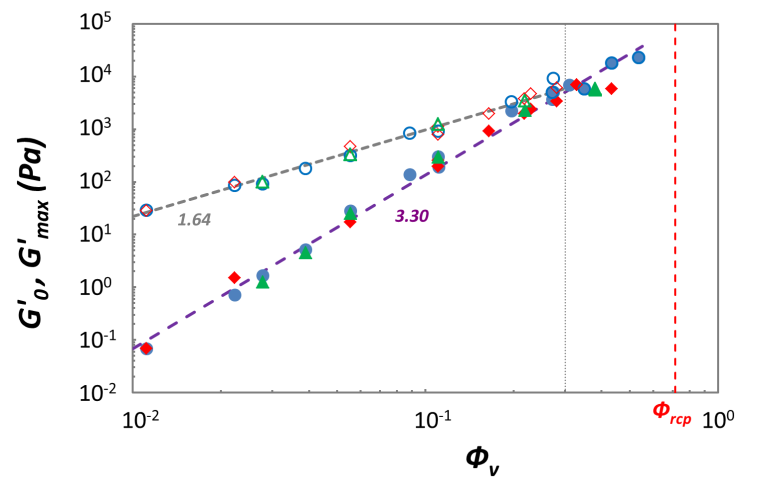}
\caption{Elastic modulus at rest $G'_{0}$  (filled symbols) and maximum elastic modulus before rupture $G’_\mathrm{max}$ (empty symbols) as a function of the rubber particles volume fraction $\phi_{v}$. The symbols and colors correspond to different Couette geometries: smooth stainless-steel (blues circles), serrated stainless steel (red diamonds) and smooth PMMA (green triangles). The red vertical dashed line shows the volume fraction at random close packing, estimated at 0.715 based on sample polydispersity (see Ref. \cite{de_Oliveira_Reis_2015}). The dashed lines represent the best power-law fits of the data, and the vertical gray line shows the limit of hardening. Inset: change in the hardening amplitude,  $G’_\mathrm{max}/G’_{0}$, with $\phi_{v}$.}
\label{fig:4}
\end{figure}

In fractal colloidal gels, the elastic modulus is given by $G'_{0}=\left(k_{0}/a\right)\phi_{v}^{\left(1+2\epsilon +d_\mathrm{b}\right)/\left(3- d_\mathrm{f}\right)}$\cite{Gisler_1999,Laxton_2007,de_Rooij_1994,de_Rooij_1993}. In this equation, $k_{0}$ is the spring constant between a pair of interacting particles, \textit{a} is the particle radius, ${\upvarepsilon}$ expresses the anisotropy of the backbone (from ${\upvarepsilon}=0$ for a straight chain to ${\upvarepsilon}=1$ for a purely isotropic chain), and $d_\mathrm{b}$ and $d_\mathrm{f}$ are the fractal dimensions of the backbone and the cluster structure respectively. If the interaction potential between particles is isotropic, aggregation leads to isotropic structures, hence ${\upvarepsilon}=1$. In this case, the network response to shear stress is controlled by bond-bending interactions between the primary particles that constitute the gel, and we obtain $(3+d_\mathrm{b})/(3-d_\mathrm{f})=3.3$. As $1\leq d_\mathrm{b}\leq 5/3$ \cite{de_Rooij_1994}, it follows that $1.59\leq d_\mathrm{f}\leq 1.79$. This is characteristic of a diffusion-limited cluster aggregation regime, for which $d_\mathrm{f}\approx 1.7-1.8$ \cite{Gauer_2009,Poon_1995,Lin_1989}, while the reaction-limited cluster aggregation regime is associated with $d_\mathrm{f}\approx 2.0-2.1$ \cite{Lin_1989,Asnaghi_1992}. 

From the values of the elastic modulus, it is also possible to estimate the spring constant between particles. Previous works on colloidal gels generally assumes that bending mechanics is dominated by contact interactions localized at the interface between two particles, which strongly depend on the physico-chemical environment \cite{Pantina_2005}. For rubber particles, the surface is covered by proteins and polar lipids, such as phospholipids and glycolipids \cite{Wadeesirisak_2017,Liengprayoon_2013}. They play a key role in latex stability and in the specific mechanical properties of dense rubber. We might therefore expect mechanical properties that strongly depend on pH or temperature, which strongly affect protein-protein or protein-phospholipid interactions.  

However, we surprisingly observed previously \cite{de_Oliveira_Reis_2015} that NRL gel mechanical properties are not affected by variations in the physico-chemical environment, for a pH ranging from 3.5 to 5.5, and temperatures ranging from 5 to \SI{40}{\celsius}. This unique behavior might originate from the nature of the rubber particles. In NRL, the core of the particles, composed of poly(cis) isoprene, is in a rubbery state at room temperature, well above the polymer glass transition that occurs around -\SI{73}{\celsius} \cite{Loadman_1985}. This is in sharp contrast with most of the colloidal gels investigated so far that are made from “hard” colloidal particles whose core is in a glassy or crystalline state. Consequently, in NRL gels, polymer chains from two different particles might locally interpenetrate at their contact point, so that the path from one center of mass to the other consists of a rubber continuum. In such a case, the mechanical properties associated with the overall inter-particle bending rigidity are not controlled by surface interactions, as usually assumed in colloidal gels, but instead by the rubber elastic modulus itself. If this holds true, the spring constant $k_{0}$ between two particle centers of mass should be equal to the dense rubber elastic modulus, $G’_\mathrm{rubber}$, times the distance between two centers of mass, here equal to two particle radii. Taking $G’_\mathrm{rubber}\approx$0.372+/-0.006  MPa \cite{Abdel-Goad2004} for the rubber elastic modulus of polyisoprene chains, a value which can vary depending on rubber properties, we find $k_{0}\approx\SI{0.37}{\pascal\metre}$. 

From our experimental results, we determine the prefactor of the power law fitted in Fig.~\ref{fig:4} as $k_{0}/a=\SI{269}{\kilo\pascal}$. With $a=\SI{0.5}{\micro\metre}$, this provides an estimate for $k_{0}$ of \SI{0.54}{\pascal\metre}, in reasonable agreement with the above estimate from rubber elastic modulus. This strongly suggests that the surface properties of the primary particles, covered in this case by proteins and phospholipids, do not play a major role in the linear mechanical properties of NRL gels, unlike what is usually observed in colloidal gels. 

 It is noteworthy that the scaling of the elastic modulus with $\phi_{v}$ holds for the full range of accessible volume fractions, while the fractal description should fail at high $\phi_{v}$. Indeed, the average cluster size at the gel point, $R_{c}$, follows $R_{C}=a\phi_{v}^{- 1/\left(3- d_\mathrm{f}\right)}$ \cite{Gisler_1999}. For volume fractions above 0.14 (for the case $d_\mathrm{b}$ = 1.59) or 0.28 (for the case $d_\mathrm{b} = 1.79$), the cluster radius drops below four particle radii, for which the fractal description is irrelevant. Still, as shown by others studies focused on internal dynamics of colloidal gels, the scaling for $G’_{0}$ remains valid, and there is no sharp mechanical signature of the transition from the dilute to the concentrated ($\phi_{v}> 0.1$) regime \cite{Romer_2014}. 

\subsection*{Hardening amplitude is volume fraction dependent}

Let us now turn to the dependence of hardening on the volume fraction of rubber particles. As shown in the inset of Fig.~\ref{fig:4}, the relative amplitude of hardening, measured as the ratio between the elastic modulus before rupture $G’_\mathrm{max}$ and the modulus at rest $G’_{0}$, decreases sharply with $\phi_{v}$. For the lowest accessible volume fraction $\phi_{v} = 0.01$, where the strain hardening amplitude is maximum, $G’_\mathrm{max}$ is about 200 times greater than $G’_{0}$ (see also Fig.~\ref{fig:3}). By contrast, for $\phi_{v}\geq 0.25$, no strain hardening is observed. This limit is in the volume fraction range previously determined for which the fractal description does not hold anymore, suggesting that the hardening phenomenon originates from the fractal nature of the particulate stress-bearing network. 

The absolute value of $G’_\mathrm{max}$ also depends on the particle volume fraction, with $G'_\mathrm{max} \phi_{v}^{1.64}$. The value of 1.64 is not compatible with a behavior dominated by bending interactions, for which ${\upvarepsilon}=1$ and the exponent of $\phi_{v}$ is given by $(3+d_\mathrm{b})/(3-d_\mathrm{f})$. Considering the uncertainty surrounding the measurement of the exponent (about +/-0.05), a precise quantification of ${\upvarepsilon}$, $d_\mathrm{b}$ and $d_\mathrm{f}$ in hardened materials is difficult. Nevertheless, this value implies that during hardening, ${\upvarepsilon}$ decreases, i.e. that structural reorganizations occurring during hardening lead to increasing backbone anisotropy. If we assume that before rupture, the mechanical response is dominated by bond-stretching, i.e. ${\upvarepsilon}\approx0$ and $d_\mathrm{b}\approx 1$ \cite{de_Rooij_1994,Wyss_2005}, we obtain $d_\mathrm{f}\approx 1.7-1.8$, which constitutes the upper limit for $d_\mathrm{f}$, suggesting that the cluster fractal dimension is not significantly modified by hardening.   

\subsection*{Characterization of the hardening effect in the nonlinear regime}

\begin{figure*}
\includegraphics[width=\textwidth]{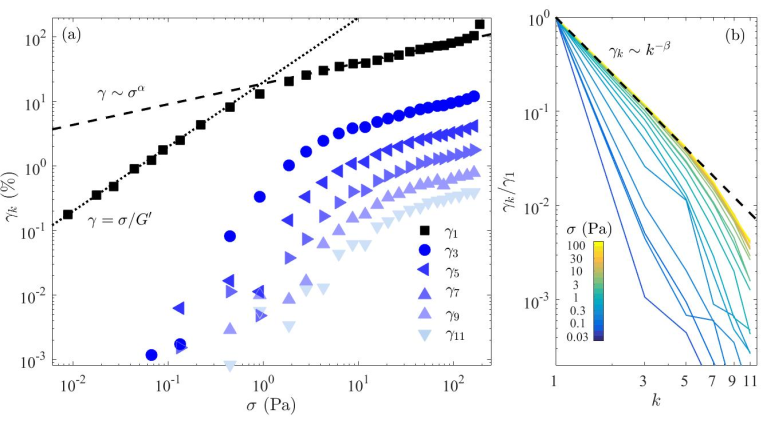}
\caption{Strain-stress relationship for $\phi_{v}$ = 0.04. (a) Change in strain harmonics $\gamma_{k}$ as a function of the applied stress amplitude $\sigma$. The dotted line is the best linear fit $\gamma_{1}={\sigma}/G’_{0}$ in the linear regime with $G’_{0}=\SI{5}{\pascal}$. The dashed line is the best power-law fit in the nonlinear regime $\gamma_{1}\sim\sigma^{\alpha}$ with $\alpha =0.32$. (b) Change in the relative intensity of odd harmonics $\gamma_{k}$ as a function of their order \textit{k}. The dashed line is $\gamma_{k}/\gamma_{1}\sim k^{-2}$.}
\label{fig:5}
\end{figure*}

To characterize the nonlinear features of the hardening phenomenon, we now focus on the Fourier components of the strain response as a function of the imposed stress. As shown previously in Fig.1, the strain response is distorted during hardening.  In Fig.~\ref{fig:5}a, the amplitude ${\gamma}_{k}$ of the harmonics of the strain response is plotted as a function of the amplitude $\sigma$ of the sinusoidal stress input.

As already described above in Fig.~\ref{fig:1}, the low-stress regime is characterized by a linear scaling of the strain fundamental amplitude $\gamma_{1}$ with the stress amplitude, $\gamma_{1}=\sigma/G’_{0}$, where $G’_{0}$ is the elastic modulus of the gel at rest (Fig.~\ref{fig:5}a). Above the onset of hardening, i.e. for $\sigma>\sigma_{0}$ and $\gamma>\gamma_0$, one has $\gamma\sim\sigma^{\alpha}$ with $\alpha\approx 0.3$ consistently with the scaling observed in Fig.~\ref{fig:1}. In the nonlinear regime, hardening is also characterized by the presence of odd harmonics, the amplitude of which ${\gamma}_{k}$ increases with $\sigma$. At high stress amplitudes, the harmonic content converges toward $\gamma_{k}/\gamma_{1}\sim k^{-\beta}$ with $\beta$ close to 2 (Fig.~\ref{fig:5}b).

\begin{figure*}
\includegraphics[width=\textwidth]{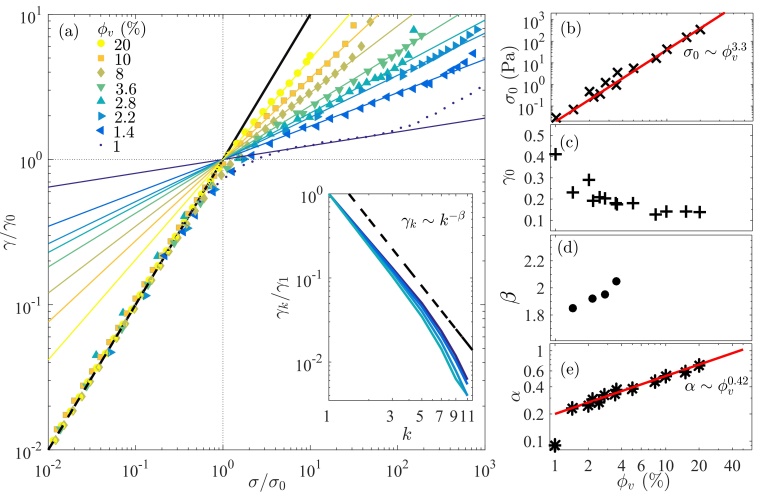}
\caption{(a) Normalized strain-stress relationships for different particle volume fractions. The black dotted line is ${\sigma}/{\sigma}_{0}={\gamma}/{\gamma}_{0}$ where ${\sigma}_{0}$ and ${\gamma}_{0}$ are the stress and strain amplitudes at the onset of hardening. Colored lines are power-law fits ${\gamma}\sim{\sigma}^\alpha$ in the hardening regime. Inset: Change in the relative intensity of the odd harmonics ${\gamma}_{k}$ as a function of their order \textit{k}. Color codes for the volume fraction. The Fourier series are computed deep into the nonlinear regime for ${\sigma}\approx 100{\sigma}_{0}$. The dashed line is ${\gamma}_{k}/{\gamma}_{1}\sim k^{-\beta}$ with $\beta=2$. (b) ${\sigma}_{0}$ as function of $\phi_{v}$. (c) ${\gamma}_{0}$ as function of $\phi_{v}$. (d) ${\beta}$ as function of $\phi_{v}$. (e) ${\alpha}$ as function of $\phi_{v}$.}
\label{fig:6}
\end{figure*}

Not surprisingly, the particle volume fraction affects the properties of the strain response (Fig.~\ref{fig:6}a). In particular, the onset stress for hardening ${\sigma}_{0}$ increases with $\phi_{v}$ (Fig.~\ref{fig:6}b), as already observed in Fig.~\ref{fig:3}. Figure 6b shows that $\sigma_{0}\sim \phi_{v}^{3.3}$ i.e. the onset stress follows the same scaling as $G’_{0}$. In other words, ${\sigma}_{0}$ is linearly related to $G’_{0}$. In contrast, the strain amplitude at the onset of hardening $\gamma_{0}$ remains constant, $\gamma_{0}\approx 0.2$, for all but the smallest volume fraction ($\phi_{v} =0.011$) for which the gels withstood a higher deformation before hardening (Fig.~\ref{fig:6}c). We conclude that above $\phi_{v} = 0.02$, the parameter that controls the hardening is the strain amplitude.

Deep into the hardening regime, the harmonic content behaves as $\gamma_{k}/\gamma_{1}\sim k^{-\beta}$ for all the volume fractions (inset of Fig.~\ref{fig:6}a). We observe a slight increase in the exponent ${\beta}$ from 1.9 to 2.1 with $\phi_{v}$ in Fig.~\ref{fig:6}d. Unfortunately, non-linear measurements at $\phi_{v}>0.05$ were out of reach in the PMMA Couette cell since the torque exceeded the limit of the rheometer. Thus, it is not clear whether the exponent ${\beta}$, and therefore the shape of the strain response, changes significantly with the particle volume fraction. Still, $\phi_{v}$ strongly affects the hardening exponent ${\alpha}$ which increases with $\phi_{v}$ as ${\alpha}\sim\phi_{v}^{0.42}$. This means that the gel hardening declines as ${\phi}_{v }$ increases. This is also seen in Fig.~\ref{fig:3} where the slope of $G’$ as a function of $\sigma$ decreases with $\phi_{v}$. These observations are in stark contrast with the reversible stress stiffening that characterizes collagen tissues \cite{Licup_2015}, for which the elastic modulus scales linearly with the stress, implying that ${\alpha}\approx 0$. In the present NRL gels, a very low value of ${\alpha}$, which also implies that the strain amplitude at rupture is very close to ${\gamma}_{0}$, is only observed for the weakest NRL gel at $\phi_{v} =0.01$. For all the other gels, hardening is observed along a significant range of deformations, typically from $\gamma_{0 }\approx 0.2$ to $\gamma_{\mathrm{max}}\approx 0.6-2$. 

\subsection*{Discussion}

\begin{figure*}
\includegraphics[width=\textwidth]{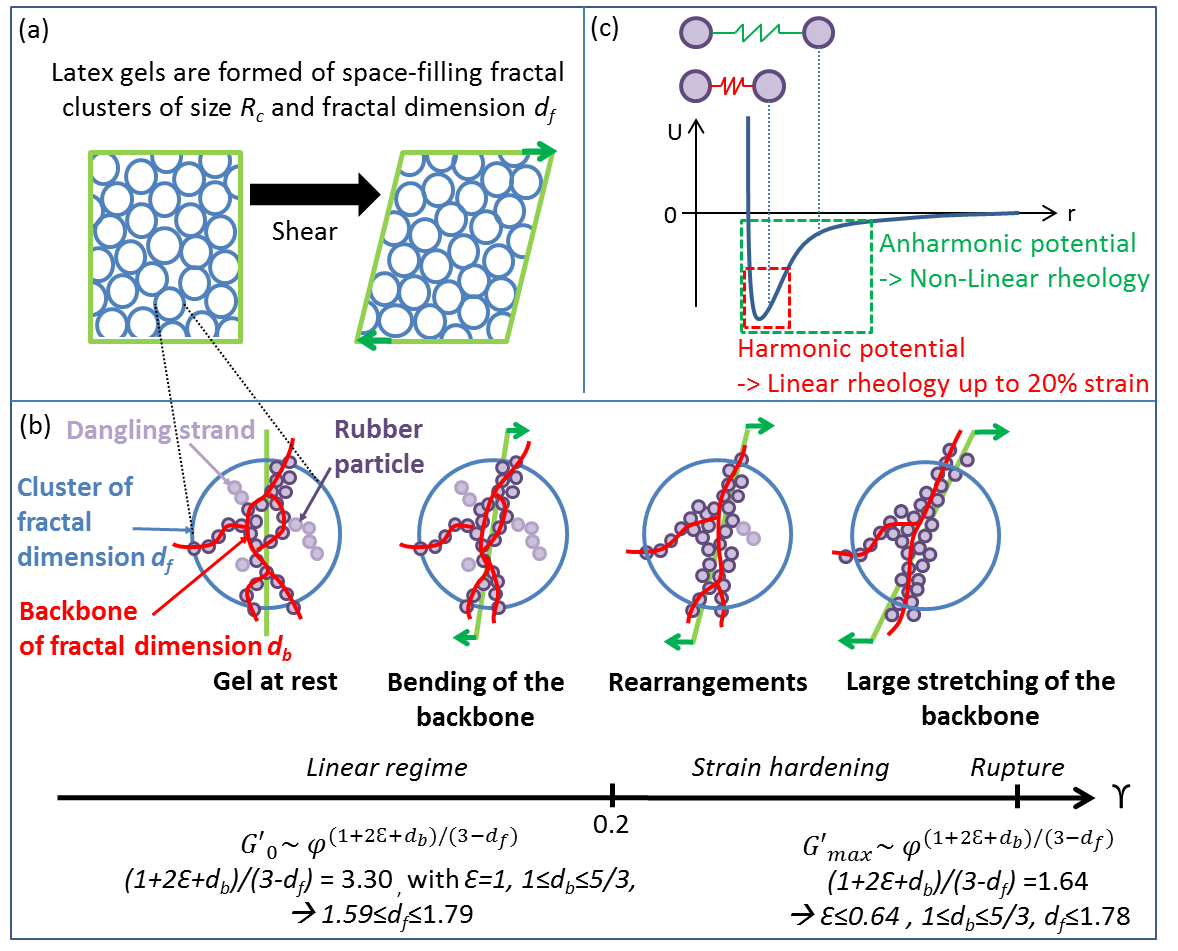}
\caption{Structural changes in NRL gels during hardening as suggested from their rheological characterization. (a) NRL gels are formed by space-filling clusters, whose size $R_{c}$ decreases with the particle volume fraction $\phi_{v}$, but whose fractal dimension $d_\mathrm{f}$ is assumed to remain constant with $\phi_{v}$. (b) In the linear regime, material deformation is locally associated with bending between particles. In the hardening regime, dangling strands are gradually incorporated into the backbone. At very large deformations, the backbone is stretched until it breaks. (c) The interparticle potential can be represented by a deep attractive well followed by a long tail: the harmonic regime is probed at small deformations while large deformations probe the anharmonic part of the potential. }
\label{fig:7}
\end{figure*}

In the following, we use the general framework of fractal gels to interpret our data and draw a simplified picture of the structural rearrangements that could lie behind strain hardening. This picture is consistent with our rheological observations but remains hypothetical in the absence of direct structural characterization of NRL gels during hardening. It should therefore be considered as a first step prior to complete system characterization. As already mentioned above, colloidal gels can be described as a dense packing of fractal clusters produced \textit{in situ} by the fractal aggregation of colloidal particles (Fig.~\ref{fig:7}a). They have been extensively studied in the literature \cite{Shih_1989,Yanez_1999}. The present results show that the spectacular irreversible hardening behavior of NRL gels depends on two control parameters: the particle volume fraction $\phi_{v}$ and the strain amplitude $\gamma$. 

We propose to describe the changes in the gel structure as a function of strain in four steps, as illustrated in Fig.~\ref{fig:7}b:  

1. At rest, for $\gamma = 0$, the gel is described as a continuous network of rubber particles. The characteristic size is the cluster size, above which the system appears homogeneous. Clusters are characterized by a fractal dimension, $d_\mathrm{f}$, which is independent of $\phi_{v}$, and by their radius, $R_{c}$, which decreases with the volume fraction \cite{Sorensen_2011}. A fraction of the particles constituting the cluster is involved in the stress-bearing backbone, which ensures the stability of the self-supported gel. This backbone is characterized by a fractal dimension, $d_\mathrm{b}$, with $d_\mathrm{b}<d_\mathrm{f}$. Another fraction of the particles is included in dangling strands that do not support stresses. At very high $\phi_{v}$, clusters are made of a very small number of particles. In such dense materials, it is likely that there are no dangling strands. However as $\phi_{v}$ decreases, $R_{c}$ increases, and the internal density of the clusters decreases. Consequently, the proportion of particles involved in dangling strands might be maximum in the gels prepared at the lowest volume fraction, and almost negligible in concentrated gels, for which the fractal theory does not hold anymore. 

2. Under small strain, for ${\gamma}< 0.2$, the response to mechanical shear stress or strain is that of a fractal gel. The elastic modulus scales with $\phi_{v}^{3.3}$, a relationship that suggests a cluster fractal dimension below 1.8, characteristic of gels formed in the diffusion-limited cluster aggregation regime. The mechanical response is governed by the resistance to bending of the interparticle bonds that are included in the cluster backbone, i.e. the particles that are involved in the stress-bearing network. In this regime, strain and stress are linearly related.

3. In the hardening regime, for ${\gamma}> 0.2$, the material opposes an increasing resistance to deformation. Hardening is however not observed for high volume fractions, which suggests that it is associated with rearrangements of the fractal structure rather than with local modifications of the particles properties themselves. Our hypothesis is that under shear, local displacements gather dangling chains on the backbone. As contacts between rubber particles lead to irreversible aggregation \cite{de_Oliveira_Reis_2015}, the backbone reinforcement is also irreversible. As the proportion of dangling strands should decrease with $\phi_{v}$, so does the maximum relative amplitude of the hardening phenomenon (inset of Fig.~\ref{fig:4}). This scenario is compatible with the results obtained from numerical simulations of colloidal gels by Colombo \textit{et al.} \textit{\cite{Colombo_2014}}, which predict that before yielding, strain can induce the formation of new bonds responsible for material stiffening. For even larger deformations, dangling strands could also interact with a neighboring backbone, creating a new connection between two adjacent backbones. These new bridges could be under tension when the overall material is at rest, but relaxed when the material is returned to the strain under which they were formed. This would explain the slight but very reproducible dip in $G’$ observed during the last cycles in Fig.~\ref{fig:2} for strains corresponding to previous hardening, a behavior that was also reported in recent simulations of soft gels \cite{Bouzid_2017}.    

4. Lastly, when all the dangling strands are included in the backbone network, stress-bearing backbones behave as stretched chains, and further deformation leads to gel fracture. 

To summarize, the above mechanism is quite unique and may be linked to two specificities of rubber particles. First, the irreversible nature of their interaction is such that when a dangling strand is captured by a neighboring backbone, it remains “stuck” to the backbone.  Second, the core of NRL particles is in a rubbery state so that particles can locally interpenetrate and interparticle bonds can sustain a very large strain before breaking. We can therefore probe large deformations and access the anharmonic part of the interparticle potential, leading to the emergence of harmonics in the rheological response (Fig.~\ref{fig:7}c). These two specificities are directly linked to the response of the stress-bearing part of the network under stress, and not to the fractal flocks interactions. They are therefore not specific to the general framework that we choose to describe our system, and would remain valid even if the local structure were better described by the so-called bond-bending percolation model \cite{Kantor_1984} or by that of a heterogeneous glass \cite{Zaccone_2009}.

\section*{{Conclusions}}

In this study, we investigated the rheological properties of Natural Rubber Latex gels. We discovered a striking hardening behavior under oscillatory shear. Although strain- and stress-controlled experiments are equivalent, we found that strain controls hardening as it appears above a characteristic strain ${\gamma}_{0}\approx 0.2$ independent of the volume fraction of rubber particles. We showed that this strain-hardening displays two unique features: (i) it is fast and irreversible and (ii) it occurs on an extended range of strains, from ${\gamma}_{0}\approx 0.2$ up to ${\gamma}\approx 0.6-2$. The relative amplitude of hardening decreases with the particle volume fraction, and ultrasonic imaging under oscillatory stress showed it to be homogeneously distributed in the gel on scales larger than a few tens of micrometers. Compared to previous reports of strain hardening in colloidal gels \cite{Gisler_1999,Pouzot_2006}, the irreversible nature of this behavior in NRL gels enables precise tuning of the material’s mechanical properties. We proposed a simplified picture based on a fractal description of the gel to explain these observations. Our hypothesis is that hardening is associated with the gradual incorporation of dangling strands into the stress-bearing backbone. In  NRL  gels, the persistence of dangling strands at rest (before deformation) could be associated with the ability of particles to partially fuse after a prolonged contact. This would explain the irreversibility of their aggregation, but would also result in oriented, non-mobile attractive interactions. Indeed, the Peclet number associated with bending a single bond can be written as $kT/\left(G'_{\mathrm{rubber}} a^{3}\right)\approx kT/\left(k_{0}a^{2}\right)\approx 10^{- 7}$ in the present case. It results that a dangling strand, once formed, should be extremely rigid and only weakly subjected to Brownian forces. 

Strain hardening/stiffening plays a key role in the mechanical behavior and properties of living cells \cite{Kollmannsberger_2011}, but it has been rarely reported in colloidal gels composed of isotropic particles. As NRL gels are composed of large particles that can be easily observed by confocal microscopy, and because their structural reorganization under strain is irreversible, a precise microstructural characterization of these materials before and after hardening is now within reach. Thus, NRL gels appear as a model tunable system for thoroughly investigating the local features of strain-stiffening in future work. Numerical simulations should also help to confirm the microscopic scenario for hardening proposed in the present work. 

\begin{acknowledgments}

This work was carried out with support from CNPq, National Council for Scientific and Technological Development -- Brazil.
\end{acknowledgments}

\bibliographystyle{apsrev4-1}
\bibliography{output}

\end{document}